\documentclass[12pt]{article}
\usepackage{amssymb}
\usepackage{amsmath}
\usepackage{graphicx}
\usepackage{bm}
\newcommand{\mathsym}[1]{{}}

\usepackage{graphicx}
\usepackage{rotating}
\usepackage{color}
\usepackage{cancel}

\newcommand{\bra}{\begin{array}}
\newcommand{\era}{\end{array}}
\newcommand{\beq}{\begin{equation}}
\newcommand{\eeq}{\end{equation}}
\newcommand{\beqar}{\begin{eqnarray}}
\newcommand{\eeqar}{\end{eqnarray}}

\newcommand{\be}{\begin{equation}}
\newcommand{\ee}{\end{equation}}
\newcommand{\bea}{\begin{eqnarray}}
\newcommand{\eea}{\end{eqnarray}}
\newcommand{\bd}{\begin{displaymath}}
\newcommand{\ed}{\end{displaymath}}

\def\BC{\bb C}
\def\_\BC{\bbi C}

\def\( {\left(}
   \def\) {\right)}
\def\[ {\left[}
\def\] {\right]}

\def\dag {{\dagger}}

\newtheorem{definition}{Definition}[section]

\newtheorem{proposition}{Proposition}[section]
\newtheorem{theorem}{Theorem}[section]

\newtheorem{corollary}{Corollary}[section]

\begin{document}

\vspace{20pt}

\begin{center}

{\LARGE \bf Three types of polynomials related to $q-$oscillator algebra
\medskip
 }
\vspace{15pt}

{\large Won Sang Chung${}^{a,*}$, Mahouton Norbert  Hounkonnou $^{b,\dag}$  and Sama  Arjika$^{b,\ddag}$ 
} 

\vspace{15pt}
{\sl ${}^a$Department of Physics and Research Institute of Natural Science,\\
 College of Natural Science,\\
 Gyeongsang National University, Jinju 660-701, Korea}\\
\vspace{15pt}
{\sl ${}^b$International Chair in Mathematical Physics and Applications\\ 
(ICMPA-UNESCO Chair), University of Abomey-Calavi,\\
072 B.P.:50, Cotonou, Rep. of Benin}\footnote{Ref-preprint:
CIPMA-MPA/015/2013}\\

\vspace{5pt}
E-mails:  {$
{}^*$mimip4444@hanmail.net,\sl $^{\dag}$norbert.hounkonnou@cipma.uac.bj, 
\sl  $^{\ddag}$rjksama2008@gmail.com}

\vspace{10pt}
\end{center}

\begin{abstract}
This work addresses a full characterization of  three new $q-$polynomials derived from the $q-$oscillator 
algebra. Related matrix elements  and
 generating functions are deduced. Further, a  connection 
between Hahn factorial  and  $q-$Gaussian polynomials is established. 
\end{abstract}

 \today
 
\section{Introduction}
\label{setiongaussian}
The $q-$deformed   Lie algebras whose applications in  the quantum field theory  and
quantum groups
 \cite{BurbanI}
possess  an 
important and useful representation theory
  in  connection to that of their classical limit algebra. The $q-$deformed  harmonic
oscillator algebra   introduced   by Arik and Coon \cite{ArikD}, and Biedenharn \cite{BiedenharnL}
   plays a similar role   as the usual boson oscillator in
   nonrelativistic quantum mechanics. This   is  why various $q-$deformed boson oscillator commutation relations
    attracted more attention during the last few years 
\cite{Balo}-\cite{Chak}.
Furthermore, quantum groups and  their representations are closely
related to the so-called $q-$calculus: $q-$numbers, $q-$factorials, $q-$differentiation, basic 
hypergeometric functions, special functions and $q-$orthogonal polynomials.
 The connection between 
special functions and group representations
was first discovered by  Cartan \cite{Cartan} in 1929. 
Vilenkin made a systematic account of 
the theory of classical special functions \cite{Vilenkin},  while  Koekoek R. et {\it al} gave a scheme on the hypergeometric 
orthogonal polynomials and their $q-$analogues \cite{ASK}.
In this scheme, all  polynomials 
are characterized by a 
set of properties:
(i) they are solutions of   difference equations, 
(ii) they are generated by  a recursion relation,
(iii) they are orthogonal with  respect to  weight function and
(iv) they obey  the Rodrigues-type formulas.

Other
polynomial families which do not obey  
 the above characteristic properties,   
do not belong to the Askey $q-$scheme. 
In this work, we deal with the study of  some 
properties of three types of polynomials: $q-$Gaussian, $q-$factorial 
and Hahn factorial polynomials. 

The paper  is organized as follows. In Section \ref{setiongaussian1}, we define 
the $q-$Gaussian polynomials. Matrix elements of the deformed   exponential functions 
 are computed  and the generating function of the $q-$Gaussian
 polynomials is deduced. The inversion formula is derived. Section \ref{setiongaussian2} is 
devoted to the $q-$factorial polynomials. Matrix elements   are computed  
 and   used to deduce some associated properties.  In Section \ref{setiongaussian3}, we recall some results on the 
Hahn calculus,   define the Hahn factorial polynomials and compute
 the  matrix elements of the new deformed exponential function 
 $E_{q,\omega}^{(\mu)}(x)$.  A connection between $q-$Gaussian 
and  Hahn factorial polynomials is established. The related inversion formula and generating
 function      are deduced in the 
Section  \ref{setiongaussian4}. We end by  some  concluding remarks in Section  \ref{setiongaussian5}.

\section{ $q-$Gaussian polynomials}
\label{setiongaussian1}
\begin{definition}
\label{qGaussians}
The $q-$Gaussian polynomials are defined as follows
\bea
\label{Gaussian}
\phi_n(x):&=&\prod_{k=0}^{n-1}(x-q^k)= (x-1)_q^n\cr
&=&\sum_{k=0}^n\left[\begin{array}{c} n \\ k \end{array}\right]_q q^{({}^k_2)}(-1)^kx^{n-k}
,\quad n\geq 1
\eea
with $\phi_0(x):=1$ and
the $q-$binomials coefficients are given by
\bea
\label{samasama:sa}
\left[\begin{array}{c} n \\ k \end{array}\right]_q:=\frac{[n]_q!}{[n-k]_q![k]_q!}=\frac{(q;q)_n}{(q;q)_{n-k}(q;q)_k}\quad \mbox{ for }\quad
0\leq k\leq n,
\eea
and zero otherwise,
\be
[n]_q:=\frac{1-q^n}{1-q},\quad [n]_q!=\prod_{k=1}^n[k]_q,
 \quad (z;q)_n:=\prod_{k=0}^{n-1}(1-zq^k),\quad (z;q)_0:=1.
\ee
\end{definition}
\begin{proposition}
The $q-$Gaussian polynomials obey   the following recursion relation
\be 
\label{qadditionm}
x\phi_n(x)=\phi_{n+1}(x)+q^n\phi_n(x), \quad \phi_0(x):=1.
\ee
\end{proposition}
{\bf Proof.} \rm Multiplying
\bea
\left[\begin{array}{c} n +1\\ k \end{array}\right]_q=\left[\begin{array}{c} n \\ k \end{array}\right]_q+q^{n+1-k}\left[\begin{array}{c} n \\ k-1 \end{array}\right]_q,
\eea
 by $q^{({}^k_2)}(-1)^kx^{n+1-k}$  and summing over $k=0,1,\cdots,n+1$, we get
\bea
\phi_{n+1}(x)&=&\sum_{k=0}^{n+1}\left[\begin{array}{c} n+1 \\ k \end{array}\right]_q q^{({}^k_2)}(-1)^kx^{n+1-k}\cr
&=&\sum_{k=0}^{n}\left[\begin{array}{c} n \\ k \end{array}\right]_q q^{({}^k_2)}(-1)^kx^{n+1-k}+q^{n+1}\sum_{k=1}^{n+1}\left[\begin{array}{c} n \\ k-1 \end{array}\right]_q q^{({}^k_2)}(-1)^kx^{n+1-k}q^{-k}\cr
&=&x\sum_{k=0}^{n}\left[\begin{array}{c} n \\ k \end{array}\right]_q q^{({}^k_2)}(-1)^kx^{n-k}+q^{n+1}\sum_{k=0}^{n }\left[\begin{array}{c} n \\ k \end{array}\right]_q q^{({}^{k+1}_{\;2})}(-1)^{k+1}x^{n-k}q^{-k-1}
\cr
&=&x\phi_n(x)-q^{n}\sum_{k=0}^{n }\left[\begin{array}{c} n \\ k \end{array}\right]_q q^{({}^{k}_{\;2})}(-1)^{k}x^{n-k} \cr
&=&x\phi_n(x)-q^{n}\phi_n(x)
\eea
which achieves the proof. $\square$
\begin{definition}
\label{def:22}
Let  $a$ and $a^\dag$ be the  operators   defined as follows:
\bea
\label{loweringandraising}
a=\frac{1-q^{x\partial x}}{x(1-q)},\qquad a^\dag =(x-1)q^{-x\partial x}
\eea
where $\partial x:=\frac{d}{dx}$ 
is the ordinary derivative and $q^{\pm x\partial x} f(x):=f(q^{\pm 1}x).$
\end{definition}
The operators $a$ and $a^\dag$  act  on the $q-$Gaussian polynomials (\ref{Gaussian}) as follows:
\begin{proposition} 
\bea
\label{samahol}
a\phi_n(x)=[n]_q\phi_{n-1}(x),\qquad a^\dag \phi_n(x)=q^{-n}\phi_{n+1}(x).
\eea
\be
\label{samaholl}
\phi_n(x)=q^{n(n-1)/2}(a^\dag)^n\cdot1.
\ee
Besides,
\be
aa^\dag  \phi_{n}(x)=q^{-n}[n+1]_q\phi_{n}(x),\qquad a^\dag a  \phi_{n}(x)=q^{-n+1}[n]_q\phi_{n}(x)
\ee
and
\be
[a,a^\dag] \phi_{n}(x)
=q^{-n}\phi_{n}(x),\qquad
[a,a^\dag]_q \phi_{n}(x)=\phi_{n}(x)
\ee
where $[A,B]:=AB-BA,\; [A,B]_q:=AB-q^{-1}BA$.
\end{proposition}
 {\bf Proof.}  See appendix A. 

Therefore,   the set of polynomials $\{\phi_n(x)|n=0,1,\cdots\}$
provides a basis for a realization of the $q-$deformed harmonic oscillator algebra
given by
\bea
aa^\dag-a^\dag a=q^{-N},\quad aa^\dag-q^{-1}a^\dag a={\bf 1},
\quad [N,a]=-a,\quad [N,a^\dag]=a^\dag
\eea
where the operator $N$ is such that
\bea
N \phi_{n}(x):=n\phi_{n}(x).
\eea
From the Definition \ref{def:22}, we deduce 
the following differential equation
\be
\Big( (x-1)q^{-x\partial x}D_x^q-[n]_{q^{-1}}\Big)\phi_n(x)=0
\ee
  for the $q-$Gaussian polynomials
where
\be
D_x^qf(x)=\frac{f(x)-f(qx)}{(1-q)x}.
\ee
In order to derive their generating function,
let us start defining the
following
$(q,\mu)-$exponential function   \cite{Elvis,flore}
\be 
\label{sama:gene1}
E_{q}^{(\mu)}(x)
=\sum_{n=0}^{+\infty}
\frac{q^{\mu n^2}}{[n]_q!}x^n,
\quad \mu \geq 0,\quad 0<q<1.
\ee
In the limit $q \to 1$,
$E_{q}^{(\mu)}(x)$ tends to the
ordinary exponential, i.e., $\lim_{q\to 1}
 E_{q}^{(\mu)}(x)= e^x,$ and  
 for some specific values of $\mu$,  it
corresponds to the standard $q-$exponentials, i.e.
  for $\mu=0$ and $\mu = 1/2$ one has
 \cite{ASK}
\bea
\label{sama:num1}
E_q^{(0)}(x/(1-q))&=& e_{q}(x) =
 \sum_{n=0}^{+\infty}\frac{1}{(q;q)_n}x^n
=\frac{1}{(x;q)_{\infty}}\\
 \label{sama:num2}
E_q^{(1/2)}(x/(1-q)) &=&E_{q}(q^{1/2}x)= \sum_{n=0}^{\infty} \frac{q^{n^2/2}}{(q;q)_n}x^n=
(-q^{1/2}x;q)_{\infty}
\eea
and  
\be
E_q^{(0)}(x)E_q^{(1/2)}(-q^{-1/2}x)=1.
\ee
Besides, let us  introduce the following operator \cite{Elvis}
\be 
\label{asabel}
\mathcal{L}^{(\mu,\nu)}(\alpha,\beta)
=E_{q}^{(\mu)}(\alpha a^\dag)
E_{q}^{(\nu)}(\beta a).
\ee
In the limit case when 
$q \to 1$, it goes into the Lie group element
$\exp(\alpha
a^\dag)\exp(\beta a).$
 Then, the matrix elements, in the
representation space spanned by the $q-$Gaussian polynomials
 $\phi_{n}(x)$,
are defined by
\be
 \mathcal{L}^{(\mu,\nu)}(\alpha,\beta)\phi_{n}(x)
 =\sum_{r=0}^{+\infty} \mathcal{L}_{n,r}^{(\mu,\nu)}
 (\alpha,\beta)
 \phi_{r}(x)
\ee
where $\mathcal{L}_{n,r}^{(\mu,\nu)}
 (\alpha,\beta)$ is  explicitly given in this work by
\be 
\label{sabel00}
\mathcal{L}_{n,r}^{(\mu,\nu)}
 (\alpha,\beta)=\beta^{n-r }q^{ \nu (n-r )^{2} }\left[\begin{array}{c} n \\ r \end{array}\right]_q\;\mathcal{U}^{(\mu,\nu)}_{r}
 \left(\alpha\beta(q-1)q^{1+2\nu(n-r)};q^{1+n-r}|q\right)
\ee 
 if  $ r\leq n,$ and
\bea
\label{sabel01}
\mathcal{L}_{n,r}^{(\mu,\nu)}
 (\alpha,\beta)=\alpha^{r-n}\,\frac{q^{\mu(r-n )^{2}  +\frac{(n-r)(n+r-1)}{2}}}{[r-n]_q!}\,
\mathcal{U}^{(\nu,\mu)}_{n}
 \left(\alpha\beta(q-1)q^{1+2\mu(r-n)};q^{1+r-n}|q\right)
\eea
  if $ n\leq r.$\\ 
The underlying polynomials  $ \mathcal{U}^{(\mu,\nu)}_n
 (x;q^{1+\theta}|q)$ are defined by the expression  
\be 
\label{sama:polyno}
 \mathcal{U}^{(\mu,\nu)}_n
 (x;q^{1+\theta}|q):=\sum_{k=0}^n
 \frac{q^{k^2(\mu+\nu)  }
 (q^{-n};q)_{k}}
 { (q^{1+\theta},q;q)_{k}} 
 x^{k}
\ee
generating  the standard
$q-$polynomials for particular
values of $\mu$ and $\nu$. Indeed,
\begin{enumerate}
\item 
For $\mu=0=\nu,$
\be
\label{sama:10}
\mathcal{U}^{(0,0)}_{n}
 (x;q^{1+\theta}|q)= {}_2\phi_{1}
\left(
\begin{array}{c}
q^{-n},0\\
q^{1+\theta}\end{array}\Big|q;x 
\right)
= p_{n}\left(xq^{-1};q^{\theta},0|q\right),
\ee
where $p_{n}(x;\gamma,\sigma|q)$ is the little
 $q-$Jacobi polynomials \cite{ASK}.
\item 
For $\mu=0,$  $\nu=\frac{1}{2}$ or vice-versa,
\be
\label{sama:101}
\mathcal{U}^{(0,1/2)}_{n}
 (x;q^{1+\theta}|q)
={}_1\phi_{1}
\left(
\begin{array}{c}
q^{-n} \\
q^{1+\theta}\end{array}\Big|q;-x q^{ \frac{1}{2}}
\right)  
= \frac{(q;q)_n}{(q^{1+\theta};q)_n}
L_{n}^{(\theta)}(-xq^{-n-\theta-1/2}|q)
\ee
where $L_{n}^{(\gamma)}(x|q)$ are the
 $q-$Laguerre polynomials \cite{ASK}.
\item 
 For $\mu=\frac{1}{2}=\nu,$
\be
\label{sama:102}
\mathcal{U}^{(1/2,1/2)}_{n}
 (x;q^{1+\theta}|q)={}_1\phi_{2}
\left(
\begin{array}{c}
q^{-n}\\
q^{1+\theta},0\end{array}\Big|q;qx
\right).
\ee
\end{enumerate}
Using the  relations (\ref{samahol}), (\ref{samaholl}) 
and the operator (\ref{asabel}),  we derive  
the generating function of the $q-$Gaussian polynomials.
\begin{theorem}
The generating function of the $q-$Gaussian  polynomials satisfies the relation
\be
\label{functiongeneratrice}
\frac{(t(1-q);q)_\infty}{( tx(1-q);q)_\infty}
=\sum_{n=0}^\infty\frac{\phi_n(x)}{[n]_q!}t^n.
\ee
\end{theorem}
{\bf Proof.}  See Appendix B. 

Note that the generating function (\ref{functiongeneratrice}) can be 
used to find different forms of formulas
appearing in this work. 
From (\ref{Gaussian}), we deduce the inversion 
formula for the $q-$Gaussian polynomials
\be
\label{masae}
x^n=\sum_{k=0}^{n}\left[\begin{array}{c} n \\ k \end{array}\right]_q \phi_k(x)
\ee
   
\begin{corollary} 
\be
{}_2\phi_{0}
\left(
\begin{array}{c}
q^{-n},x^{-1}\\
-\end{array}\Big|q;xq^n
\right)=x^n
\ee
and
\be
\sum_{j=0}^{n}\left[\begin{array}{c} n \\ j \end{array}\right]_qq^{({}^j_2)}(-1)^j{}_2\phi_{0}
\left(
\begin{array}{c}
q^{-n+j},
0\\
-\end{array}\Big|q;x q^{n-j}
\right)=x^n.
\ee
\end{corollary}
%
From the Definition \ref{qGaussians}, the  
$q-$Gaussian polynomials can be also determined under the form
\be
\phi_n (x) = E_q^{(1/2)} \big( - q^{-1/2} D_x^q   \big)  x ^n.
\ee
Let  $|\psi\rangle :=\sum_{n=0}^{\infty} c_n(x)\phi_n(x) $ be the eigenvector of 
the position operator $X:=a^\dag+a $ with the 
the eigenvalue $x$, i.e.,
\be
\label{sama:recur}
X|\psi\rangle=x|\psi\rangle. 
\ee
Since (\ref{samahol}) is satisfied, equating coefficients of 
the polynomials $\phi_n(x)$ in  both the sides of (\ref{sama:recur}), we 
obtain the following three-term recursion relation for the 
coefficients $c_n(x)$:
\be
xc_n(x)=[n+1]_qc_{n+1}(x)+q^{1-n}c_{n-1}(x),\quad n\geq 1
\ee
with $c_0(x):=1.$ 
Since $[n]_q\neq q^{1-n},$  the position operator is not symmetric. Immediatly, one can see  that
\be
c_{2n}(0)=(-1)^n\frac{q^{n(1-n)}}{[2n]_q!!},\qquad c_{2n+1}(0)=0.
\ee 
As matter of illustration, we compute the first five coefficients as follows:
\bea
c_1(x)&=&x\\
c_2(x)&=&\frac{1}{[2]_q}(x^2-1)\\
c_3(x)&=&\frac{1}{[3]_q!}(x^3- x(1+q^{-1}[2]_q))\\
c_4(x)&=&\frac{1}{[4]_q!}\left(x^4-x^2(1+q^{-1}[2]_q+q^{-2}[3]_q) +q^{-2}[3]_q\right)\\
c_5(x)&=&\frac{1}{[5]_q!}\left(x^5-x^3(1+q^{-1}[2]_q+q^{-2}[3]_q+
q^{-3}[4]_q) +x(q^{-2}[3]_q + q^{-3}[4]_q+q^{-4}[2]_q[4]_q)\right).\cr
&&
\eea
\section{$q-$Factorial polynomials}
\label{setiongaussian2}
\begin{definition}
The $q-$factorial polynomials are defined as follows:
\bea
\hat{\phi}_n(x)=\frac{\Gamma_q(x+1)}{\Gamma_q(x+1-n)}=
\prod_{k=0}^{n-1}[x-k]_q,\quad n\geq 1
\eea
with $\hat{\phi}_0(x):=1.$
\end{definition}
The $q-$factorial polynomials behave as ordinary monomials 
under the action of the operators
\be
\label{samaqa}
\hat{a}^\dag=[x]_qe^{-\partial x},\qquad \hat{a}=q^{-x-1}(e^{\partial x}-1)
\ee
where $e^{\pm \partial x} f(x):=f(x\pm 1)$.
Indeed, $a$ and $a^\dag$ are called step operators when they 
appear in the $q-$deformed quantum
theory and they satisfy the following relation.
\begin{proposition}
\be
\label{msa}
\hat{a}^\dag\hat{\phi}_n(x)=\hat{\phi}_{n+1}(x),\qquad  
\hat{a} \hat{\phi}_n(x)=q^{-n}[n]_q\hat{\phi}_{n-1}(x).
\ee
\end{proposition}
{\bf Proof.} By acting $\hat{a}^\dag$ on the $q-$factorial 
polynomials $\hat{\phi}_n(x)$, we have
\bea
\hat{a}^\dag\hat{\phi}_n(x)&=&[x]_q\hat{\phi}_n(x-1)\cr
&=&[x]_q\prod_{k=0}^{n-1}[x-(1+k)]_q\cr
&=&[x]_q[x-1]_q\cdots[x- n ]_q\cr
&=&\hat{\phi}_{n+1}(x).
\eea
In the same way, 
by acting $\hat{a}$ on the $q-$factorial 
polynomials $\hat{\phi}_n(x)$, one gets
\bea
\hat{a} \hat{\phi}_n(x)&=& q^{-x-1}(\hat{\phi}_n(x+1)-\hat{\phi}_n(x))\cr
&=&q^{-x-1}\Big( \prod_{k=0}^{n-1}[x+1-k]_q-\prod_{k=0}^{n-1}[x-k]_q\Big)\cr
&=&q^{-x-1}\Big([x+1]_q-[x-n+1]_q\Big) \prod_{k=0}^{n-2}[x-k]_q\cr
&=&q^{-n}[n]_q\hat{\phi}_{n-1}(x)
\eea
which achieves the proof. $\square$

Since (\ref{msa}) holds, we have
\be
\hat{a}\hat{a}^\dag  \hat{\phi}_{n}(x)=q^{-n-1}[n+1]_q\hat{\phi}_{n}(x),
\qquad \hat{a}^\dag \hat{a} \hat{\phi}_{n}(x)=q^{-n}[n]_q\hat{\phi}_{n}(x)
\ee
and
\be
[\hat{a},\hat{a}^\dag] \hat{\phi}_{n}(x)=q^{-n-1}\hat{\phi}_{n}(x),\quad [\hat{a},\hat{a}^\dag]_q \hat{\phi}_{n}(x)
=q^{-1}\hat{\phi}_{n}(x).
\ee
Therefore,   the set of polynomials $\{\hat{\phi}_n(x)|n=0,1,\cdots\}$
provides a basis for a realization of the $q-$deformed harmonic oscillator algebra
given by
\bea
\hat{a}\hat{a}^\dag-\hat{a}^\dag \hat{a}=q^{-N-1},\quad 
\hat{a}\hat{a}^\dag-q^{-1}\hat{a}^\dag \hat{a}=q^{-1},\quad 
[N,\hat{a}]=-\hat{a},\quad [N,\hat{a}^\dag]=\hat{a}^\dag.
\eea
Following the previous development  for the $q-$Gaussian 
polynomials, the  introduction of the operator 
\be 
\label{samaasmaa}
\hat{\mathcal{L}}^{(\mu,\nu)}(\alpha,\beta)
=E_{q}^{(\mu)}(\alpha \hat{a}^\dag)
E_{q}^{(\nu)}(\beta\hat{a}),
\ee
leads to 
the
 matrix elements, in the
representation space spanned by the $q-$factorial polynomials
 $\hat{\phi}_{n}(x)$ 
  defined by
\bea
 \hat{\mathcal{L}}^{(\mu,\nu)}(\alpha,\beta)\hat{\phi}_{n}(x)
=
 \sum_{r=0}^{+\infty}\hat{\mathcal{L}}_{n,r}^{(\mu,\nu)}
 (\alpha,\beta)
 \hat{\phi}_r(x)
\eea
where $\hat{\mathcal{L}}_{n,r}^{(\mu,\nu)}
 (\alpha,\beta)$ are here explicitly given by
\bea 
 \hat{\mathcal{L}}_{n,r}^{(\mu,\nu)}(\alpha,\beta)=
 \beta^{n-r}q^{\nu(n-r)^{2}+\frac{(r-n)(n+r+1)}{2}}
 {n
\atopwithdelims[] r}_q {\mathcal{U}}^{(\mu,\nu)}_{r}
 \left(\alpha\beta(q-1)q^{2\nu(n-r)};q^{1+n-r}|q\right)
 \label{samamat1}
\eea
if $r\leq n,$ and
\be 
\hat{\mathcal{L}}_{n,r}^{(\mu,\nu)}(\alpha,\beta)=
\frac{ \alpha ^{r-n}
q^{\mu(r-n)^2}}{[r-n]_q!} \;{ \mathcal{U}}^{(\nu,\mu)}_{n}
 \left(\alpha\beta(q-1)q^{2\mu(r-n)};q^{1+r-n}|q\right)
 \label{samamat2}
\ee
if $n\leq r.$ 

Therefore, the following statement   holds.
\begin{theorem}
The generating function  for the $q-$factorial  polynomials is defined by
\be
\label{sfunctiongeneratrice}
{}_2\phi_{0}
\left(
\begin{array}{c}
q^{-x},0\\
-\end{array}\Big|q;tq^x
\right)=\sum_{n=0}^\infty\frac{\hat{\phi}_n(x)}{[n]_q!}t^n.
\ee
or 
\be
{}_1\phi_{0}
\left(
\begin{array}{c}
q^{-x}\\
-\end{array}\Big|q;-t q^{x }
\right)=\sum_{n=0}^\infty\frac{q^{({}^n_2)}\hat{\phi}_n(x)}{[n]_q!}t^n.
\ee
\end{theorem}
{\bf Proof.} From the   definition of the matrix 
elements (\ref{samaasmaa}), we have
\be 
\label{samarie}
 \hat{\mathcal{L}}^{(\mu,0)}(\alpha,0).1
=
 E_{q}^{(\mu)}(\alpha \hat{a}^\dag).1=\sum_{n=0}^\infty\frac{q^{\mu n^2}\alpha^n}{[n]_q!}\hat{\phi}_n(x)
\ee 
\begin{itemize}
\item 
If $\mu=0$, we arrive at the generating function of the $q-$factorial polynomials
\be 
 E_{q}^{(0)}(\alpha \hat{a}^\dag).1=\sum_{n=0}^\infty\frac{\hat{\phi}_n(x)}{[n]_q!}\alpha^n
\ee  
By using the identity
\be
\prod_{k=0}^{n-1}[x-k]=(-1)^nq^{nx-({}^n_2)}\frac{(q^{-x};q)_n}{(1-q)^n}
\ee
we get
\be 
 \sum_{n=0}^\infty\frac{\hat{\phi}_n(x)}{[n]_q!}\alpha^n
=\sum_{n=0}^\infty\frac{(-1)^nq^{- ({}^n_2)}
(q^{-x};q)_n}{(q;q)_n}(\alpha q^x)^n={}_2\phi_{0}
\left(
\begin{array}{c}
q^{-x},0\\
-\end{array}\Big|q;\alpha q^x
\right).
\ee  
\item 
If $\mu=1/2$, (\ref{samarie}) takes the form
\bea
\label{sama:e} 
 \hat{\mathcal{L}}^{(1/2,0)}(\alpha,0).1
&=&\sum_{n=0}^\infty\frac{ (q^{-x};q)_n}{(q;q)_n}(-\alpha q^{x+1/2})^n\cr
&=&{}_1\phi_{0}
\left(
\begin{array}{c}
q^{-x}\\
-\end{array}\Big|q;-\alpha q^{x+1/2}
\right).
\eea 
\end{itemize}
The rest is  achieved by setting $t=\alpha q^{1/2}$ on 
the right hand-side of (\ref{sama:e}). $\square$

\section{ Hahn calculus: Hahn factorial polynomials}
\label{setiongaussian3}
Quantum difference operators are receiving an increasing interest in 
applied mathematics and  theoretical physics because of
 their numerous applications \cite{Almeida}-\cite{Kac}. 
Further, the
quantum calculus generates  the ordinary derivative by a difference operator, which allows us
to treat sets of non-differentiable functions. Since Jackson \cite{Jackson} introduced the first expression
of difference operator, called the Jackson derivative, several 
expressions of the difference operator appeared. Among them, the most famous one is Hahn's difference operator \cite{Hahn}, which
has two deformation parameters  $\omega$ and $q$. Hahn's operator reduces to Jackson's $q-$derivative
when the parameter $\omega$ goes to $0$.

\subsection{   Hahn's calculus}
\begin{definition}
 The Hahn's derivative is defined as follows
\be
D_{q,\omega}f(x)=\left\{\begin{array}{ll}
\frac{f(qx+\omega)-f(x)}{(q-1)x+\omega}&\quad x\neq\omega_0\\
f'(x)&\quad x=\omega_0,
\end{array}\right.
\ee
where  $\omega_0=\frac{\omega}{1-q}.$
\end{definition}
\begin{theorem}
The Hahn's derivative satisfies the following deformed Leibniz rule:
\bea
D_{q,\omega}(f(x) g(x))&=&(D_{q,\omega}f(x))g(x)+f(qx+\omega)D_{q,\omega}g(x)\\
D_{q,\omega}\left(\frac{f(x)}{ g(x)}\right)&=&\frac{(D_{q,\omega}f(x))g(x)-f(x)D_{q,\omega}g(x)}{g(x)g(qx+\omega)}.
\eea
\end{theorem}
The proof is straighforward. See Appendix C.

The Hahn integral is introduced as the inverse operation of the Hahn derivative, i.e.,
\begin{theorem}
 The Hahn integral is defined by
\be
\int_{\omega_0}^xf(x')d_{q,\omega}x'=((1-q)x-\omega)
\sum_{k=0}^\infty q^k f(xq^k+\omega [k]_q).
\ee
\end{theorem}
{\bf Proof.} If $D_{q,\omega}F(x)=f(x),$ we have
\bea
\label{sama;sa}
&&F(x)-F(qx+\omega)=((1-q)x-\omega)f(x)\\
&&F(qx+\omega)-F(q^2x+[2]_q\omega) =((1-q)qx-q\omega)f(qx+\omega)\\
&&\vdots\cr
&&F(q^nx+\omega)-F(q^{n+1}x+[n+1]_q\omega) =((1-q)q^nx-q^n\omega)f(q^nx+[n]_q\omega)\label{sama;saa}
\eea
Summing (\ref{sama;sa}) to (\ref{sama;saa}), we arrive at
\be 
 F(x)-F(q^{n+1}x+[n+1]_q\omega) =((1-q) x- \omega)\sum_{k=0}^nq^kf(q^kx+[k]_q\omega).
\ee 
When $n\to\infty,$ the latter expression takes the form
\be 
 F(x)-F(  \omega_0) =\int_{\omega_0}^xf(x')d_{q,\omega}x'=((1-q) x- \omega)\sum_{k=0}^\infty q^kf(q^kx+[k]_q\omega)
\ee 
what achieves the proof. $\square$
\subsection{ Hahn factorial polynomials}
\begin{definition}
The  Hahn factorial polynomials are defined by
\be
\label{sama:hahn}
\dot{\phi}_n(x)=\prod_{k=0}^{n-1}(x-[k]_q\omega),\quad n\geq 1
\ee
with $\dot{\phi}_0(x):=1, $ satisfying the following recursion relation
\be 
x\dot{\phi}_n(x)=\dot{\phi}_{n+1}(x)+\omega[n]_q\dot{\phi}_n(x),
\quad n\geq 1,\quad \dot{\phi}_0(x):=1.
\ee
\end{definition}

%
\begin{theorem}
Let $\dot{a}$ and $\dot{a}^\dag$ be the  operators defined as follows:
\bea
\label{sama:loweringandraising}
\dot{a}=D_{q,\omega},\qquad \dot{a}^\dag =xq^{-x\partial x-\omega\partial \omega}e^{-\omega\partial x}
\eea
where $e^{\pm \omega\partial x} f(x):=f(x\pm \omega)$. Then, 
\bea
\label{samahold}
\dot{a}\dot{\phi}_n(x)=[n]_q\dot{\phi}_{n-1}(x),\qquad \dot{a}^\dag\dot{\phi}_n(x)=q^{-n}\dot{\phi}_{n+1}(x).
\eea
and
\be
\label{sabelma}
\dot{\phi}_n(x)=q^{n(n-1)/2}(\dot{a}^\dag)^n\cdot1.
\ee
\end{theorem}
{\bf Proof.}
\bea
\dot{a}\dot{\phi}_n(x)=D_{q,\omega}\dot{\phi}_n(x)&=&\frac{\dot{\phi}_n(qx+\omega)-\dot{\phi}_n(x)}{(q-1)x+\omega}\cr
&=&\frac{1}{(q-1)x+\omega}\Big(\prod_{k=0}^{n-1}(xq+\omega-[k]_q\omega)-\prod_{k=0}^{n-1}(x-[k]_q\omega)\Big)\cr
&=&\frac{1}{(q-1)x+\omega}\Big((qx+\omega)q^n\prod_{k=1}^{n-1}(x-[k-1]_q\omega)-\prod_{k=0}^{n-1}(x-[k]_q\omega)\Big)\cr
&=&\frac{1}{(q-1)x+\omega}((q-1)[n]_qx+\omega[n]_q)\prod_{k=0}^{n-2}(x-[k]_q\omega)
\cr
&=&[n]_q\dot{\phi}_{n-1}(x).
\eea
In the same way, we have
\bea
\dot{a}^\dag\dot{\phi}_n(x)&=&xq^{-x\partial x-\omega\partial \omega}e^{-\omega\partial x}\dot{\phi}_n(x)\cr
&=&x\dot{\phi}_n(q^{-1}(x-\omega))\cr
&=&xq^{-n}\prod_{k=0}^{n-1}(x-\omega-q[k]_q\omega)\cr
&=&xq^{-n}\prod_{k=0}^{n-1}(x-[k+1]_q\omega)\cr
&=&q^{-n}\dot{\phi}_{n+1}(x)
\eea
which achieves the proof. $\square$

Since (\ref{samahold}) holds, we get
\be
\dot{a} \dot{a}^\dag  \dot{\phi}_{n}(x)=q^{-n}[n+1]_q\dot{\phi}_{n}(x),\qquad \dot{a}^\dag \dot{a}\dot{\phi}_{n}(x)=q^{-n+1}[n]_q\dot{\phi}_{n}(x)
\ee
and
\be 
[\dot{a},\dot{a}^\dag] \dot{\phi}_{n}(x)=q^{-n}\dot{\phi}_{n}(x),\qquad
[\dot{a},\dot{a}^\dag]_q\dot{\phi}_{n}(x)
=\dot{\phi}_{n}(x).
\ee
Therefore,   the set of polynomials $\{\dot{\phi}_n(x)|n=0,1,\cdots\}$
provides a basis for a realization of the $q-$deformed harmonic oscillator algebra
given by
\be 
\dot{a}\dot{a}^\dag-\dot{a}^\dag \dot{a}=q^{-N},\qquad \dot{a}\dot{a}^\dag-q^{-1}\dot{a}^\dag \dot{a}={\bf 1},\quad [N,\dot{a}]=-\dot{a},\quad [N,\dot{a}^\dag]=\dot{a}^\dag.
\ee 
\begin{definition}
The  Hahn exponential function $e_{q,\omega}(x)$ is defined as
\be
\label{sama:samasa}
D_{q,\omega}e_{q,\omega}(x)=e_{q,\omega}(x).
\ee
\end{definition}
Thus, we have the following.
\begin{theorem}
\be
e_{q,\omega}(x)=\frac{e_{q,\omega}(\omega_0)}{\prod_{k=0}^\infty(1+q^k((q-1)x+\omega))}.
\ee
\end{theorem}
{\bf Proof.} Since (\ref{sama:samasa}) is satisfied, we have
\be 
\frac{e_{q,\omega}(qx+\omega)-e_{q,\omega}(x)}{(q-1)x+\omega}=e_{q,\omega}(x)
\ee 
equivalent to
\bea
e_{q,\omega}(x)&=&\frac{e_{q,\omega}(qx+\omega)}{1+(q-1)x+\omega}\cr
&=&\frac{e_{q,\omega}(q^2x+(1+q)\omega)}{(1+(q-1)x+\omega)(1+q(q-1)x+q\omega))}\cr
&\vdots&\cr
&=&\frac{e_{q,\omega}(q^sx+(1+q+q^2+q^{s-1})\omega)}{\prod_{k=0}^{s-1}(1+q^k((q-1)x+\omega))}\cr
&=&\frac{e_{q,\omega}(q^sx+[s]_q\omega)}{\prod_{k=0}^{s-1}(1+q^k((q-1)x+\omega))}.
\eea
When $s\to\infty,\,|q|< 1,$ the latter expression takes the form
\be 
e_{q,\omega}(x) =\frac{e_{q,\omega}(\omega_0)}{\prod_{k=0}^\infty(1+q^k((q-1)x+\omega))}
\ee 
which achieves the proof. $\square$

In the limit when $\omega_0\to 0$, we recover the well known exponential function $E_q^{(0)}(x).$

The definition of the matrix elements for the Hahn factorial polynomials
requires the construction of the 
$(q,\omega,\mu)-$exponential function in the form:
\be 
\label{samasagene1}
E_{q,\omega}^{(\mu)}(x)
=\sum_{n=0}^{+\infty}
\frac{q^{\mu n^{2}}}{(q;q)_n}((1-q)x-\omega)^n,
\quad \mu \geq 0,\quad 0<q<1,
\ee
giving, for $\omega=0,$   the $(q,\mu)-$exponential function 
$E_{q}^{(\mu)}((1-q)x)$ investigated  in   \cite{Elvis,flore}. For $\mu=0$,  
\be 
\label{samasage}
e_{q,\omega}(x)=e_{q,\omega}(\omega_0)E_{q,\omega}^{(0)}(x),
\ee
while in the limit $q \to 1,\;\omega=0$,
$E_{q,\omega}^{(\mu)}(x)$ tends to the
ordinary exponential: $\lim_{q\to 1}
 E_{q,0}^{(\mu)}(x)= e^x$.
For $\omega=0=\mu$ and $\omega=0,\;\mu = 1/2$ one has
 \cite{ASK}
\bea
\label{sama:num1}
E_{q,0}^{(0)}(x/(1-q))&=& e_{q}(x) =
 \sum_{n=0}^{+\infty}\frac{1}{(q;q)_{n}}x^{n}
=\frac{1}{(x;q)_{\infty}}\\
 \label{sama:num2}
E_{q,0}^{(1/2)}(x/(1-q)) &=&E_{q}(q^{-1/2}x)= \sum_{n=0}^{\infty} \frac{q^{n(n-1)/2}}{(q;q)_n}x^n=
(-q^{-1/2}x;q)_{\infty}
\eea
  satisfying
\be
E_{q,0}^{(0)}(x)E_{q,0}^{(1/2)}(-q^{1/2}x)=1.
\ee
 Introduce the previous  operator 
\be
\tilde{\mathcal{L}}^{(\mu,\nu)}(\alpha,\beta)
=E_{q,\alpha \omega \dot{a}^\dag}^{(\mu)}(\alpha  \dot{a}^\dag)
E_{q,\beta \omega  \dot{a}}^{(\nu)}(\beta  \dot{a}) 
\ee
going into the Lie group element
$\exp(\alpha
 \dot{a}^\dag)\exp(\beta \dot{a})$ in the limit
$\omega=0,\;q \to 1.$
Their matrix elements, in the
representation space spanned by the Hahn factorial polynomials
 $\dot{\phi}_{n}(x)$,
are defined by
\bea
\label{sama:matrix} 
 \tilde{\mathcal{L}}^{(\mu,\nu)}(\alpha,\beta)\dot{\phi}_{n}(x)
=\sum_{r=0}^{+\infty} \tilde{\mathcal{L}}_{n,r}^{(\mu,\nu)}
 (\alpha,\beta)
 \dot{\phi}_{r}(x)
\eea
where $\tilde{\mathcal{L}}_{n,r}^{(\mu,\nu)}
 (\alpha,\beta)$  is explicitly given by
\bea 
\label{sabel}
\tilde{\mathcal{L}}_{n,r}^{(\mu,\nu)}(\alpha,\beta)&=&
 (\beta(1-\omega_0))^{n-r}q^{\nu(n-r)^{2}}
 {n
\atopwithdelims[] r}_q \cr
&\times&{\mathcal{U}}^{(\mu,\nu)}_{r}
 \left(\alpha\beta(q-1)(1+\omega_0)^2q^{1+2\nu(n-r)};q^{1+n-r}|q\right) 
 \label{samasamat1}
\eea
if $r\leq n,$ and
\bea
\label{sabel2}
\tilde{\mathcal{L}}_{n,r}^{(\mu,\nu)}(\alpha,\beta)&=&
\frac{ (\alpha(1-\omega_0)) ^{r-n}
q^{\mu(r-n)^2+\frac{(n-r)(n+r-1)}{2}}}{[r-n]_q!}\cr
&\times&{\mathcal{U}}^{(\nu,\mu)}_{n}
 \left(\alpha\beta(q-1)(1+\omega_0)^2q^{1+2\mu(r-n)};q^{1+r-n}|q\right) 
 \label{samasamat2}
\eea
 if $n\leq r$.

In the limit when $\omega_0$ goes to $0$, the matrix elements (\ref{sabel}) 
and (\ref{sabel2}) are reduced  to (\ref{sabel00}) 
and (\ref{sabel01}), respectively.

\section{Connection between the  Hahn factorial and the $q$-Gaussian polynomials}
\label{setiongaussian4}
In this section, we establish a connection between the Hahn factorial 
and the $q-$Gaussian polynomials. The inversion formula and generating 
function related to the  Hahn factorial polynomials are   given. \\
Let us start with an  alternative definition of the Hahn factorial 
polynomials (\ref{sama:hahn}) as follows:
\begin{definition}
\be
\label{sama:Hfact}
\dot{\phi}_n(x):=\sum_{k=0}^n\left[\begin{array}{c} n \\ k \end{array}\right]_q q^{({}^k_2)}\omega_0^k(x-\omega_0)^{n-k}
,\quad n\geq 1
\ee
with $\dot{\phi}_0(x):=1$ and $\omega_0=\omega/(1-q),$ satisfying the recursion relation
\be
\label{sama:recurrence}
(x-\omega_0)\dot{\phi}_n(x)=\dot{\phi}_{n+1}(x)-\omega_0\, q^n\dot{\phi}_n(x).
\ee
\end{definition}
\begin{proposition}
The  connection formula between the Hahn factorial polynomials 
$\dot{\phi}_n(x)$ (\ref{sama:Hfact}) and the $q-$Gaussian polynomials $ \phi_n(x)$ is given by
\be
\label{sama:connection}
\dot{\phi}_n(x)=(-1)^n\omega_0^n\,{\phi}_n(1-x/\omega_0).
\ee
\end{proposition}
{\bf Proof.} Immediately, one can  see  that
\be
x-\omega[k]_q
=-\omega_0\Big(1-\frac{x}{\omega_0}-q^k\Big).
\ee
 Then, the relation (\ref{sama:hahn}) takes the form
\bea
\dot{\phi}_n(x)
&=&\prod_{k=0}^{n-1}-\omega_0\Big(1-\frac{x}{\omega_0}-q^k\Big)\cr
&=&(-1)^n\omega_0^n\,{\phi}_n(1-x/\omega_0).
\eea
The rest of the proof is achieved by combining (\ref{qadditionm}) and (\ref{sama:connection}). $\square$

Since  (\ref{masae}) is valid, the  inversion formula for the  Hahn factorial polynomials is
\be
\label{sam:sama:}
 (x-\omega_0)^n=\sum_{k=0}^{n}\left[\begin{array}{c} n \\ k \end{array}\right]_q (-1)^{n-k}\omega_0^{n-k}\; \dot{\phi}_k(x).
\ee
\begin{theorem}
The generating function  for the Hahn factorial  polynomials   is defined by
\be
\label{functiongeneratricesq}
\frac{(-t\omega;q)_\infty}{( -t((q-1)x+\omega);q)_\infty}
=\sum_{n=0}^\infty\frac{\dot{\phi}_n(x)}{[n]_q!}t^n.
\ee
\end{theorem}
{\bf Proof.} From the   definition of the matrix elements  (\ref{sama:matrix}) and the relation (\ref{sabelma}), we have
\be 
  \tilde{\mathcal{L}}^{(\mu,0)}(\alpha,0).1
=
 E_{q,\alpha\omega \dot{a}^\dag}^{(\mu)}(\alpha  \dot{a}^\dag)\cdot1=\sum_{n=0}^\infty\frac{q^{\mu n^2-({}^n_2)}\dot{\phi}_n(x)}{(q;q)_n} (\alpha(1-q-\omega) )^n.
\ee 
If $\mu=1/2$, we arrive at the generating function of the  Hahn factorial polynomials
\bea 
\label{sqamaytsummationnew}
  E_{q,\alpha\omega \dot{a}^\dag}^{(1/2)}(\alpha  \dot{a}^\dag)\cdot1&=&\sum_{n=0}^\infty\frac{\dot{\phi}_n(x)}{(q;q)_n} (\alpha(1-q-\omega)q^{1/2})^n\cr
&=&\sum_{n=0}^\infty\frac{\phi_n(1-x/\omega_0)}{[n]_q!} (\alpha\omega_0(\omega_0-1)q^{1/2})^n.
\eea  
 The proof is achieved by   using 
  (\ref{functiongeneratrice}) and setting $t=\alpha (1-\omega_0)q^{1/2}$ on the
right-hand side of (\ref{sqamaytsummationnew}). $\square$

\section{Concluding remarks}
\label{setiongaussian5}
In this work, we have studied three types of $q-$polynomials related to the $q-$oscillator 
algebra. Matrix elements of each family of
polynomials are computed  and the 
associated generating functions are deduced. Finally, a  connection 
between Hahn factorial  and the $q-$Gaussian polynomials is established. 

\section*{Acknowledgements}
MNH and SA acknowledge   the Abdus Salam International
Centre for Theoretical Physics (ICTP, Trieste, Italy) for its support through the
Office of External Activities (OEA) - \mbox{Prj-15}. The ICMPA
is also in partnership with
the Daniel Iagolnitzer Foundation (DIF), France.
\section*{Appendix A}
From the definition, we have
\bea
a\phi_n(x)&=&\frac{1-q^{x\partial x}}{x(1-q)}\prod_{k=0}^{n-1}(x-q^k)\cr
&=&\frac{1}{x(1-q)}\Big(\prod_{k=0}^{n-1}(x-q^k)-\prod_{k=0}^{n-1}(qx-q^k)\Big)
\cr
&=&\frac{1}{x(1-q)}\Big(\prod_{k=0}^{n-1}(x-q^k)-q^n\prod_{k=0}^{n-1}(x-q^{k-1})\Big)
\cr
&=&\frac{1}{x(1-q)}\Big((x-q^{n-1})-q^n(x-q^{-1})\Big)\prod_{k=0}^{n-2}(x-q^{k})
\cr
&=&\frac{1-q^n}{1-q}\phi_{n-1}(x).
\eea
Similarly, 
\bea
a^\dag\phi_n(x)&=&(x-1)q^{-x\partial x}\prod_{k=0}^{n-1}(x-q^k)\cr
&=&(x-1) \prod_{k=0}^{n-1}(q^{-1}x-q^k)\cr
&=&q^{-n}(x-1) \prod_{k=1}^{n}( x-q^{k})
\cr
&=&q^{-n}  \prod_{k=0}^{n }( x-q^{k})
\cr
&=& q^{-n}\phi_{n+1}(x)
\eea
which achieves the proof. $\square$
\section*{Appendix B}
From the   definition of the matrix elements  (\ref{asabel}), we have
\be 
  {\mathcal{L}}^{(\mu,0)}(\alpha,0).1
=
 E_{q}^{(\mu)}(\alpha  {a}^\dag).1=\sum_{n=0}^\infty\frac{q^{\mu n^2-({}^n_2)}\alpha^n}{[n]_q!} {\phi}_n(x)
\ee 
If $\mu=1/2$, we arrive at the generating function of the $q-$Gaussian polynomials
\be 
\label{sama:summationnew}
 E_{q}^{(1/2)}(\alpha a^\dag).1=\sum_{n=0}^\infty\sum_{k=0}^n\frac{ (-1)^kq^{({}^k_2)}}{[n-k]_q![k]_q!} (q^{1/2}\alpha)^nx^{n-k}
\ee  
By introducing the new summation index $m=n-k$ on the right-hand side of  (\ref{sama:summationnew}), one
obtains
\be
\label{sama:summationnews}
 {\mathcal{L}}^{(1/2,0)}(\alpha,0).1=e_q (q^{1/2}\alpha x(1-q))E_q(-q^{1/2}\alpha(1-q)).
\ee
The proof is achieved by taking $t=\alpha q^{1/2}$  on the
right-hand side of (\ref{sama:summationnews}). $\square$
\section*{Appendix C}
{\bf Proof.} Using the definition, 
\bea
D_{q,\omega}(f(x) g(x))&=&\frac{(f(qx+\omega) g(qx+\omega)-(f(x) g(x)}{(q-1)x+\omega}\cr
&=&(f(qx+\omega)\frac{ g(qx+\omega)- g(x)}{(q-1)x+\omega}+g(x)\frac{f(qx+\omega)-f(x) }{(q-1)x+\omega}\cr
&=&f(qx+\omega)D_{q,\omega}g(x)+g(x)D_{q,\omega}f(x).
\eea
In the same way,
\bea
D_{q,\omega}\left(\frac{f(x)}{ g(x)}\right)&=&\frac{\frac{f(qx+\omega)}{ g(qx+\omega)}-\frac{f(x)}{ g(x)}}{(q-1)x+\omega}\cr
&=&\frac{f(qx+\omega)g(x)-g(qx+\omega)f(x)}{((q-1)x+\omega)g(x)g(qx+\omega)}\cr
&=&g(x)\frac{f(qx+\omega)-f(x)}{((q-1)x+\omega)g(x)g(qx+\omega)}-f(x)\frac{g(qx+\omega)-g(x)}{((q-1)x+\omega)g(x)g(qx+\omega)}\cr
&=&\frac{(D_{q,\omega}f(x))g(x)-f(x)D_{q,\omega}g(x)}{g(x)g(qx+\omega)}
\eea
which achieves the proof. $\square$

\end{document}